\documentclass[aps,reprint,amsmath,amssymb,groupaddress,superscriptaddress]{revtex4-2}
\usepackage{graphicx}
\usepackage[utf8]{inputenc}
\usepackage{amsmath}
\usepackage{subfigure}
\usepackage{amsfonts}
\usepackage{amssymb}
\usepackage{bm}
\usepackage{dcolumn}
\usepackage{color}
\usepackage[utf8]{inputenc}
\usepackage[T1]{fontenc}
\usepackage{booktabs, array, mathptmx, float, tabularx, booktabs, lipsum, amsmath, multirow,mathtools}
\graphicspath{{figs/}{figsgaoerb/}} 
\usepackage{soul}
\usepackage{CJK}
\usepackage{xcolor}
\usepackage{mathtools}
\usepackage{lipsum}
\usepackage{comment}

\usepackage{xcolor}
\usepackage[colorlinks,citecolor=blue,linkcolor=red,anchorcolor=blue,urlcolor=blue]{hyperref}

\begin{document}

\title{Revisiting Endo-reversible Carnot engine: Extending the Yvon engine}
\author{Xiu-Hua Zhao}
\affiliation{School of Physics and Astronomy, Beijing Normal University, Beijing, 100875, China}

\author{Yu-Han Ma}
 \email{yhma@bnu.edu.cn}
\affiliation{School of Physics and Astronomy, Beijing Normal University, Beijing, 100875, China}
\affiliation{Graduate School of China Academy of Engineering Physics, No. 10 Xibeiwang East Road, Haidian District, Beijing, 100193, China}

\date{\today}
\begin{abstract}
A famous paper 
[\href{https://doi.org/10.1119/1.10023}{Am. J. Phys. 43, 22 (1975)}] unveiled the efficiency at maximum power (EMP) of the endo-reversible Carnot heat engine, now commonly referred to as the Curzon-Ahlborn (CA) engine, pioneering finite-time thermodynamics. Historically, despite the significance of the CA engine, similar findings had emerged at an earlier time, such as the Yvon engine proposed by J. Yvon in 1955 sharing the exact same EMP. However, the special setup of the Yvon engine has circumscribed its broader influence. This paper extends the Yvon engine model to achieve a level of generality comparable to that of the CA engine. A rigorous comparison reveals that the extended Yvon engine and CA engine represent the steady-state and cyclic forms of the endo-reversible Carnot heat engine, respectively, and are equivalent. Our work provides a pedagogical example for the teaching of thermodynamics and engineering thermodynamics, given that the simple and lucid derivation of the extended Yvon engine helps students initiate their understanding of non-equilibrium thermodynamics.

\end{abstract}

\maketitle
\section{Introduction}

The seminal paper by Curzon and Ahlborn~\cite{curzonEfficiencyCarnotEngine1975} published in 1975 
is a landmark work in finite-time thermodynamics~\cite{andresen2011current}. By analyzing an endoreversible Carnot engine with finite temperature differences between the working substance and the reservoirs, the authors optimized the power output of the engine and obtained the efficiency at maximum power (EMP), which is a more practical efficiency bound than the Carnot efficiency. The EMP formula they derived, 
\begin{equation}
\label{eq-CA-efficiency}
    \eta_{\mathrm{CA}}=1-\sqrt{T_c/T_h}
\end{equation}
with $T_c$ ($T_h$) the temperature of the cold (hot) reservoir, is commonly referred to as Curzon-Ahlborn (CA) efficiency. This result has become a cornerstone in the study of finite-time heat engines~\cite{tu2012recent,leffReversibleIrreversibleHeat2018}.

In fact, Curzon and Ahlborn’s work was not the first to explore the performance of finite-time heat engines or derive the CA efficiency~\cite{bejanModelsPowerPlants1996,moreauCarnotPrincipleIts2015,ouerdaneContinuityBoundaryConditions2015}. As early as 1955, Yvon presented the same formula as Eq.~\eqref{eq-CA-efficiency} while investigating the optimization of actual power plants modeled as endoreversible engines~\cite{yvonSaclayReactorAcquired1955}. Two years later, Chambadal~\cite{chambadalCentralesNucleaires1957} and Novikov~\cite{novikovEfficiencyAtomicPower1957} presented discussions quite similar to that of Yvon and reached the same conclusion. Yvon's approach, though yielding the same EMP expression, differed from Curzon and Ahlborn's in two key aspects: i) a finite temperature difference was assumed only between the working substance and the hot reservoir, with no temperature difference on the cold side, and ii) instead of explicitly considering cycle time, finite heat flux and work rate were directly used to characterize the finite-time nature of practical engines. These distinctions render the relationship between the Yvon and CA engines less straightforward, obfuscating the grounds for their having the same EMP.

The simplicity of Yvon’s optimization procedure inspires us to extend his model to encompass more general finite-time engines. The extended Yvon engine presented in this work provides a unified and straightforward framework for studying endoreversible Carnot engines, bridging the gap between Yvon’s original analysis and the broader scope of Curzon and Ahlborn’s work. Specifically, we relax the assumption of equal temperatures on the cold side in the Yvon engine, and optimize the work rate (power) with respect to the endoreversible temperatures of the working substance under the endoreversible condition. The derived EMP aligns with the CA efficiency, yet the optimization process is significantly more concise than that of the CA engine. Moreover, our approach naturally reproduces the trade-off relation between power and efficiency beyond the maximum-power regime~\cite{chenEffectHeatTransfer1989}, a topic of considerable interest in contemporary thermodynamic research~\cite{tuAbstractModelsHeat2021,maFinitetimeThermodynamicsJourney2024}. We also provide a rigorous comparison between the two models to demonstrate their equivalence in describing the endoreversible Carnot heat engine.


\begin{figure*}[!tbh]
    \centering
    \includegraphics[width=0.48\linewidth]{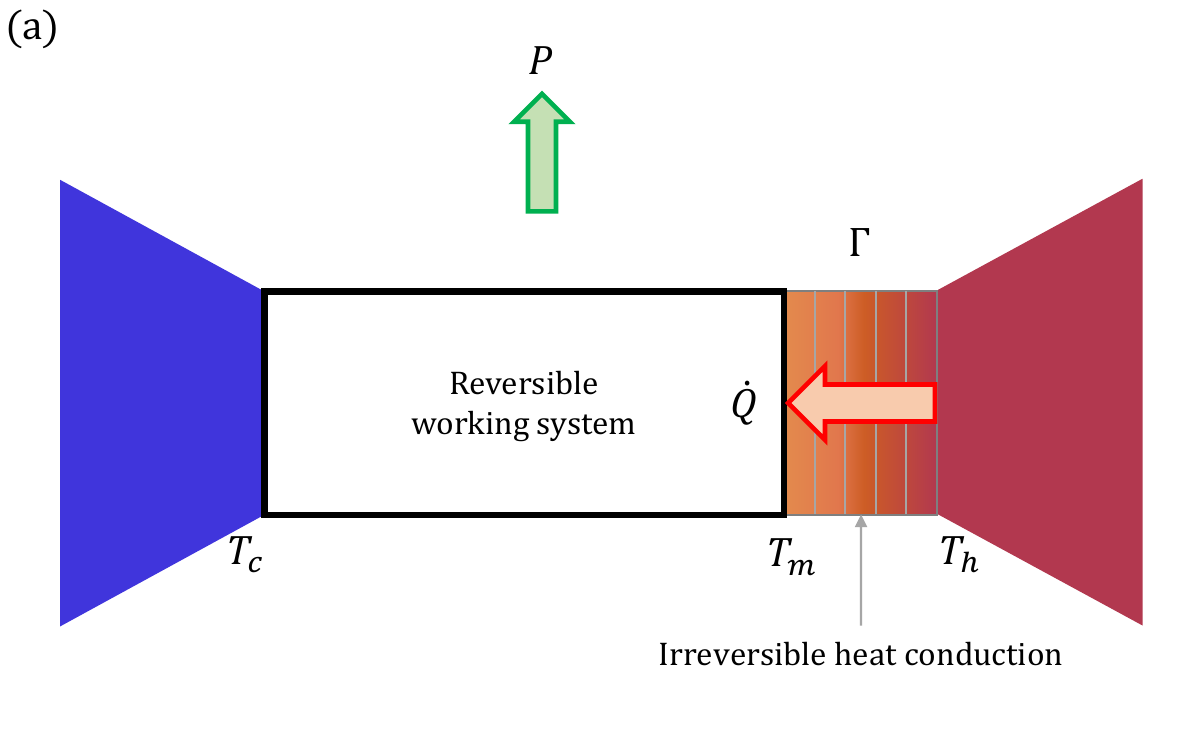}
    \includegraphics[width=0.48\linewidth]{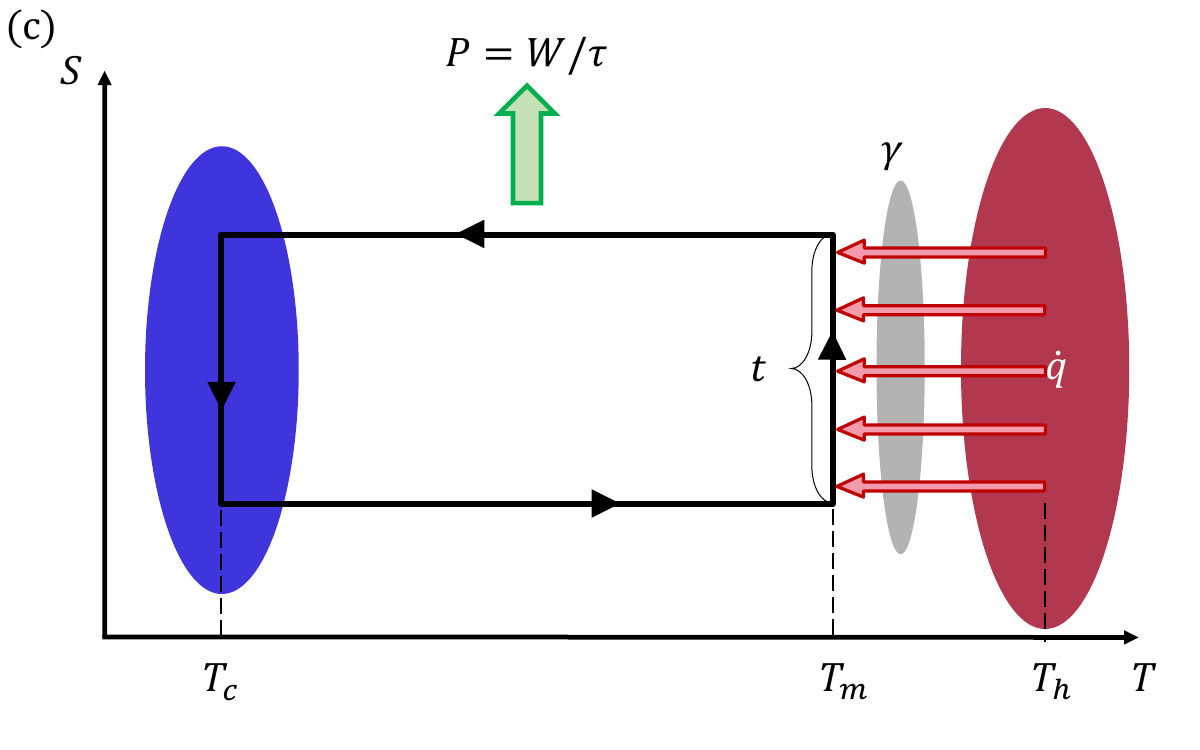}
    \includegraphics[width=0.48\linewidth]{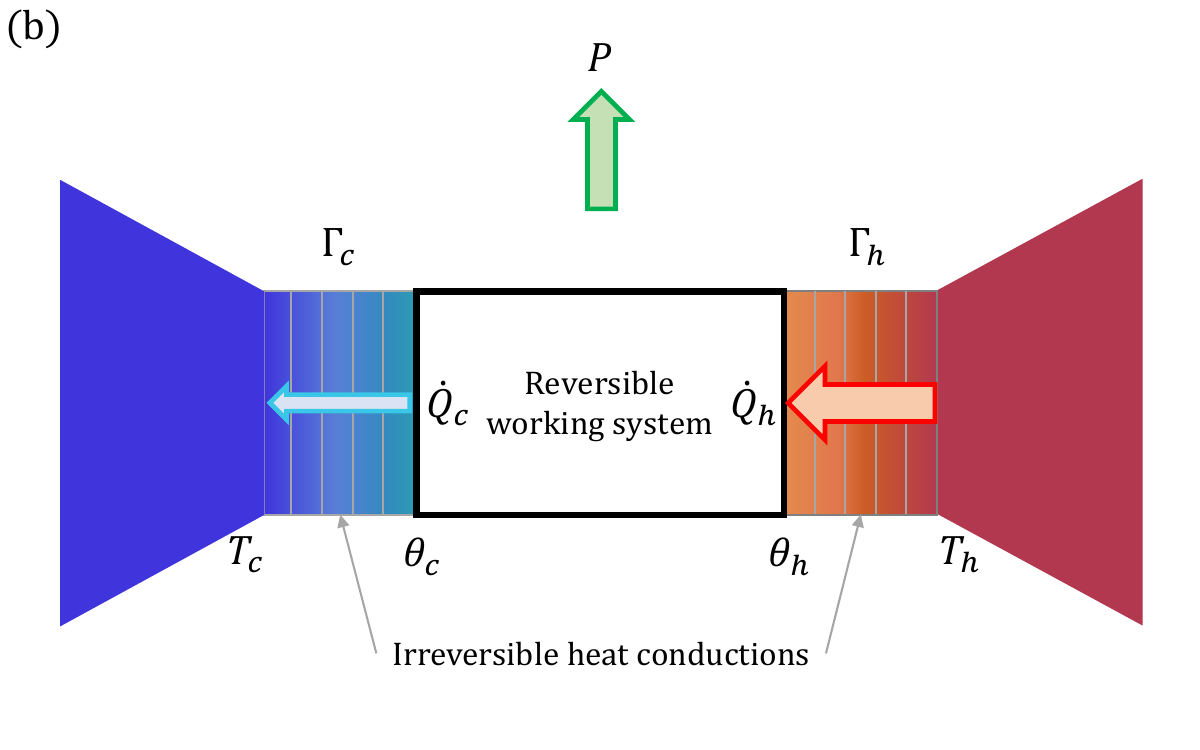}
    \includegraphics[width=0.48\linewidth]{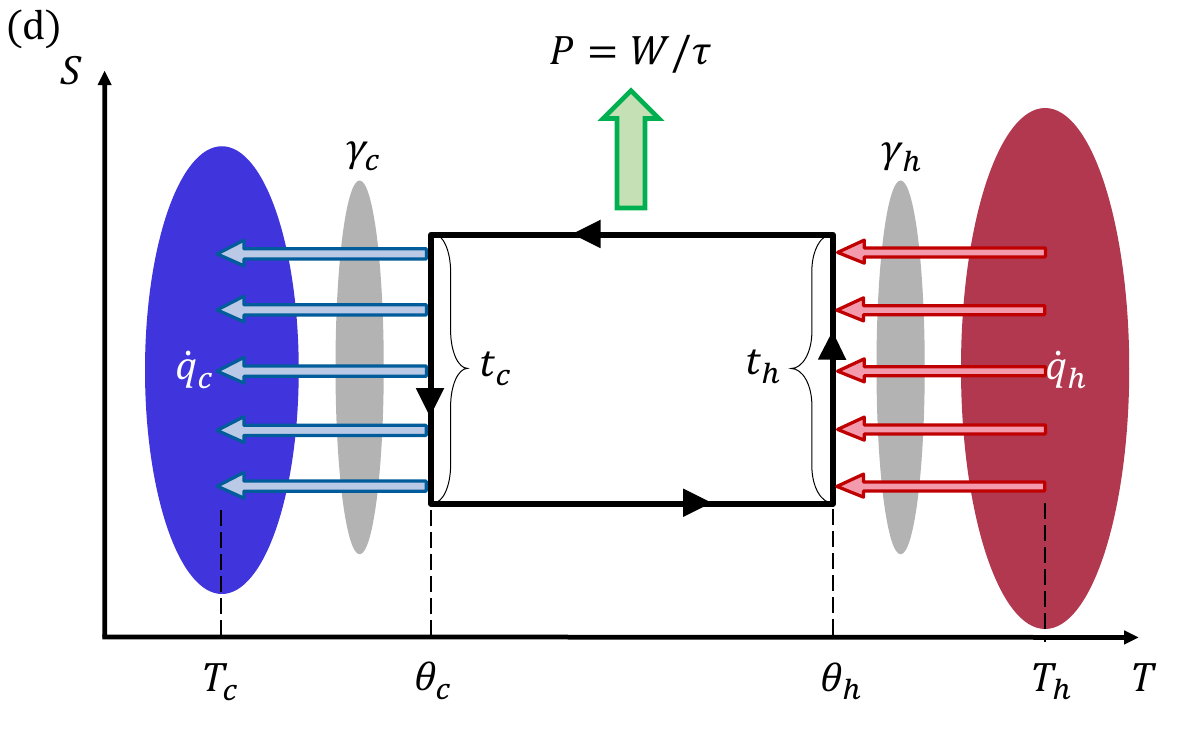}
    \caption{Steady-state [(a) and (b)] and cyclic [(c) and (d)] endoreversible heat engines. (a) In the original Yvon engine, finite heat flux (denoted as $\dot{Q}$) occurs only at the high-temperature end, where there is a temperature difference between the working substance and the hot reservoir. (b) The extended Yvon engine introduces temperature differences, and thus heat fluxes, between the working substance and both the hot and cold reservoirs. (c) and (d) show the Entropy ($S$)-Temperature ($T$) diagrams of the endoreversible Carnot engine cycles with finite heat fluxes (denoted as $\dot{q}$) along the high-temperature and both isothermal branches, respectively.}
    \label{fig:cycles}
\end{figure*}

\section{\label{Sec:Yvon}Yvon engine}

We start with reviewing Yvon's pioneering treatment on the finite-time Carnot heat engine~\cite{moreauCarnotPrincipleIts2015,yvonSaclayReactorAcquired1955}. Yvon's report~\cite{yvonSaclayReactorAcquired1955} at the 1955 United Nations Conference on the Peaceful Uses of Atomic Energy offered a simple solution on maximizing the output power of the steam engine in nuclear reactors. Specifically, as shown in Fig.~\ref{fig:cycles} (a), the fluid, at temperature $T_h$ from the reactor core, serves as the hot reservoir transferring heat to the hot end of the working substance in the engine, where the temperature of this end is maintained at $T_m<T_h$. The Yvon engine generates power through the turbine shafts and isothermally releases heat to the condenser where the working substance and the condenser are at the same lowest temperature $T_c<T_m$. Assume that the heat flux from the fluid to the working substance obeys Newton's law of cooling, namely,
\begin{equation}
\label{eq-heat-flux-Yvon}
    \dot{Q} = \Gamma(T_h - T_m),
\end{equation}
where $\Gamma$ is a constant that depends on the thermal conductivity and area of the wall separating the working substance from the hot fluid, and the overdot notation stands for rates. The heat engine efficiency is defined as $\eta\equiv P/\dot{Q}$, where $P$ is the output power of the engine. With the assumption that the working substance undergoes a reversible transformation between $T_m$ and $T_c$, the efficiency of the engine reads
\begin{equation}
    \eta^{\mathrm{(Y)}} = 1 - T_c/T_m,
\end{equation}
which is now known as the endoreversible assumption~\cite{rubinOptimalConfigurationClass1979}. Hence, the power of the Yvon engine follows as
\begin{equation}\label{PY}
    \begin{aligned}
        P^{\mathrm{(Y)}}= \eta^{\mathrm{(Y)}}\dot{Q}=&\Gamma \left( T_h+T_c-T_m-\frac{T_hT_c}{T_m} \right)\\
        \le&\Gamma\left(\sqrt{T_h}-\sqrt{T_c}\right)^2,
    \end{aligned}
\end{equation}
where the AM-GM inequality, $(a+b)/2\ge\sqrt{ab}$, for non-negative $a$ and $b$ has been applied. Associated with the third equality in the above equation, the maximum power is achieved when 
\begin{equation}
    T_m =T_m^*\equiv \sqrt{T_h T_c},
\end{equation}
and the corresponding efficiency at maximum power $\eta^{\mathrm{(Y)}}|_{T_m=T^*_m}=\eta_{\mathrm{CA}}$ is easily checked.

\section{\label{Sec:extended Yvon}Extended Yvon engine}

In the Yvon engine, heat flux is considered only when the working substance absorbs heat from the hot reservoir. In this section, we extend the model to include heat flux on the cold side, resulting from the temperature difference between the working substance and the cold reservoir, as illustrated in  Fig.~\ref{fig:cycles}(b). This extended Yvon engine is consistent with the CA engine [Fig.~\ref{fig:cycles}(d)], though it uses a different representation method based on energy fluxes rather than examining each branch of a thermodynamic cycle.

The heat flux from the hot reservoir at temperature $T_h$ to the working substance at temperature $\theta_h<T_h$ follows
\begin{equation}
\label{eq-heat-flux-hot}
    \dot{Q}_h=\Gamma_h\left( T_h-\theta_h \right),
\end{equation}
where the constant $\Gamma_h$ characterizes the heat conductivity during heat absorption. Similarly, the heat flux from the working substance, whose temperature decreases to $\theta_c<\theta_h$, to the cold reservoir at temperature $T_c<\theta_c$ is
\begin{equation}
\label{eq-heat-flux-cold}
    \dot{Q}_c=\Gamma_c\left(\theta_c-T_c\right),
\end{equation}
with $\Gamma_c$ being a constant during heat release. As a result of the energy conservation law, the engine's output power is
\begin{equation}
\label{eq-power}
    P=\dot{Q}_h-\dot{Q}_c=\eta\dot{Q}_h.
\end{equation}
In this case, the endoreversible assumption reads
\begin{equation}
\label{eq-entropy-variation}
    \frac{\dot{Q}_h}{\theta_h}=\frac{\dot{Q}_c}{\theta_c},
\end{equation}
and correspondingly, $\eta =1-\theta_c/\theta_h$ is the endoreversible efficiency. 
It follows from Eqs.~(\ref{eq-heat-flux-hot}), (\ref{eq-heat-flux-cold}) and (\ref{eq-entropy-variation}) that $\theta_h$ can be expressed with the temperature ratio $\theta_h/\theta_c$ as
\begin{equation}\label{relation}
\theta_h=\frac{\Gamma_h T_h}{\Gamma_h+\Gamma_c}+\frac{\Gamma_c T_c}{\Gamma_h+\Gamma_c}\frac{\theta_h}{\theta_c},
\end{equation}
substituting which into Eq.~(\ref{eq-power}), we obtain
\begin{equation}\label{PYvon}
        P=\frac{\Gamma_h \Gamma_c T_h}{\Gamma_h+\Gamma_c}\left(1+\frac{T_c}{T_h}-\frac{\theta_c}{\theta_h}-\frac{T_c}{T_h}\frac{\theta_h}{\theta_c}\right).
\end{equation}
Similar to Eq.~\eqref{PY}, by utilizing the AM-GM inequality, it is straightforward to find that, for given $\Gamma_{h,c}$ and $T_{h,c}$,  
\begin{equation}\label{PYvonmax}
    P\le\frac{\Gamma_h \Gamma_c T_h}{\Gamma_h+\Gamma_c}\left(1-\sqrt{\frac{T_c}{T_h}}\right)^2\equiv P_{\mathrm{max}},
\end{equation}
where the maximum power $P_{\mathrm{max}}$ is achieved with the optimal endoreversible temperatures $\theta_h^*$ and $\theta_c^*$ which satisfy
\begin{equation}\label{optimalT}
    \frac{\theta_c^*}{\theta_h^*}=\sqrt{\frac{T_c}{T_h}}.
\end{equation}
Consequently, the EMP of the engine $\eta_{\rm{MP}}=1-\theta^*_c/\theta^*_h=\eta_{\rm{CA}}$. As $\Gamma_h/\Gamma_c\rightarrow0$, along with Eqs.~(\ref{relation}) and (\ref{optimalT}), we have $\theta^*_h=\sqrt{T_h T_c}=T^*_m$ and $\theta^*_c=T_c$, thus recovering the original Yvon engine.
\begin{figure}[!tbh]
    \centering
    \includegraphics[width=0.95\linewidth]{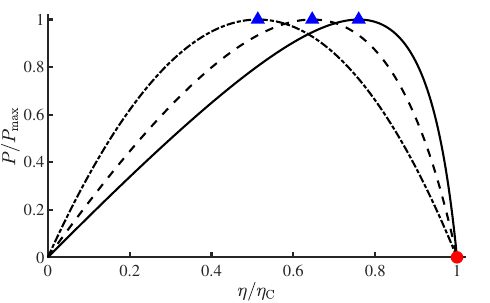}
    \caption{Trade-off relations between power and efficiency for the CA engine with $\eta_{\mathrm{C}}=0.1$ (dash-dotted curve), $\eta_{\mathrm{C}}=0.7$ (dashed curve), and $\eta_{\mathrm{C}}=0.9$ (solid curve). The triangles and circle mark the maximum power and maximum efficiency, respectively.}
    \label{fig:tradeoff}
\end{figure}

Furthermore, in Eq.~(\ref{PYvon}), replacing the temperature ratios with efficiencies as $T_c/T_h=1-\eta_{\mathrm{C}}$ and $\theta_c/\theta_h=1-\eta$ yields
\begin{equation}\label{Petarelation}
    P=\frac{T_h}{\Gamma_h^{-1}+\Gamma_c^{-1}}\frac{\eta\left(\eta_{\mathrm{C}}-\eta\right)}{\left(1-\eta\right)},
\end{equation}
which is illustrated in Fig.~\ref{fig:tradeoff} for $\eta_{\mathrm{C}}=0.1,0.7,0.9$ with normalized axes. This relation was first obtained by Chen and Yan~\cite{chenEffectHeatTransfer1989} through solving $(\partial P/\partial \theta_h)_{\eta}=0$ for the cyclic CA engine, determining the efficiency at arbitrary power or the power at arbitrary efficiency of the endoreversible Carnot engine. The explicit dependence of power on efficiency, which provides a comprehensive optimization regime for thermal machines, is now referred to as the trade-off or constraint relation between power and efficiency~\cite{holubecEfficiencyMaximumPower2015,shiraishi2016universal,maUniversalConstraintEfficiency2018,yuanOptimizingThermodynamicCycles2022,zhaiExperimentalTestPowerefficiency2023}, and constitutes one of the key focuses within the realm of finite-time thermodynamics~\cite{ma2024unified}. For more details and related progress on this issue, please refer to Ref.~\cite{maFinitetimeThermodynamicsJourney2024} along with the references encompassed therein.

\section{\label{Sec:comparison}Comparison of Curzon-Ahlborn and extended Yvon engines}

Both the Yvon engine and its extended version are steady-state heat engines, while the CA engine adopts a cyclic representation, taking into account the duration of each thermodynamic process in the cycle. Although the previous section has demonstrated the consistency of the EMP obtained by the extended Yvon and CA engines, their strict correspondence requires further clarification. In this section, we will establish the strict correspondence between the parameters of the two models. As the cyclic counterpart of the extended Yvon engine [Fig.~\ref{fig:cycles}(b)], the CA engine is depicted in Fig.~\ref{fig:cycles}(d) via the Entropy-temperature ($S-T$) diagram. Meanwhile, Fig.~\ref{fig:cycles}(c) represents the limiting case of the CA heat engine where the temperature difference at the low-temperature end approaches zero, corresponding to the original Yvon engine in Fig.~\ref{fig:cycles}(a).

In Curzon and Ahlborn's original derivation~\cite{curzonEfficiencyCarnotEngine1975}, engine power is expressed not through energy fluxes but as the total work $W=Q_h-Q_c$ divided by the duration $\tau=\xi(t_h+t_c)$ of an endoreversible Carnot cycle, namely,
\begin{equation}
\label{eq-power-CA}
    P^{\mathrm{(CA)}}=\frac{Q_h-Q_c}{\xi \left(t_h+t_c\right)}.
\end{equation}
Here, $t_h$ and $t_c$ represent the durations of the heat absorption and release processes [see Fig.~\ref{fig:cycles}(d)], respectively, while the time taken to complete the adiabatic transitions, $(\xi-1)(t_h+t_c)$, is assumed to be proportional to the duration of isothermal processes with $\xi>1$. The exchanged heat, $Q_{h(c)}=\int\dot{q}_{h(c)}dt$, in the isothermal processes is
\begin{equation}
   Q_h=\gamma_h(T_h-\theta_h)t_h,~Q_c=\gamma_c(\theta_c-T_c)t_c,  \label{heatCA} 
\end{equation}
where $T_{h,c}$ and $\theta_{h,c}$ have the same meanings as in the extended Yvon model. However, the heat transfer coefficients $\gamma_{h,c}$ are different from $\Gamma_{h,c}$, as will be discussed later. As the result of the endoreversible assumption and the cyclic condition, the entropy variation of the working substance in a cycle satisfies
\begin{equation}
\label{eq-entropy-variation-CA}
    \Delta S=\frac{Q_h}{\theta_h}-\frac{Q_c}{\theta_c}=0.
\end{equation}
Consequently, this cyclic engine's efficiency, $\eta\equiv(Q_h-Q_c)/Q_h=1-\theta_c/\theta_h$. Comparing the first equality in Eq.~(\ref{eq-power}) with Eq.~(\ref{eq-power-CA}) and Eq.~(\ref{eq-entropy-variation}) with Eq.~(\ref{eq-entropy-variation-CA}), we obtain the relation between $\gamma_{h,c}$ and $\Gamma_{h,c}$ as
\begin{equation}
\Gamma_{h(c)}=\frac{t_{h(c)}}{\xi\left(t_h+t_c\right)}\gamma_{h(c)}.\label{Gamma}
\end{equation}
This equation shows that the heat transfer coefficients in the steady-state engine differ from those in the cyclic engine by a time-proportional factor, which is determined by the ratio of the corresponding process duration to the total cycle duration. 

According to Eq.~\eqref{Gamma}, the overall coefficient appearing in the upper bound of Eq.~(\ref{PYvonmax}) satisfies
\begin{equation}\label{gamma}
        \frac{\Gamma_h \Gamma_c }{\Gamma_h+\Gamma_c}=\frac{\gamma_h \gamma_c}{\xi\left(1+t_c/t_h\right)\left(\gamma_h t_h/t_c +\gamma_c\right)}
        \le  \frac{\gamma_h \gamma_c}{\xi \left( \sqrt{\gamma_h}+\sqrt{\gamma_c} \right)^2},
\end{equation}
where $\gamma_h t_h/t_c+\gamma_c t_c/t_h\ge 2\sqrt{\gamma_h \gamma_c}$ has been used, and the equality is saturated with the optimal time ratio 
\begin{equation}
    \frac{t_h^*}{t_c^*}=\sqrt{\frac{\gamma_c}{\gamma_h}}.
\end{equation} 
This indicates that when the process durations are taken into account, cyclic engines have an additional parameter to be optimized, i.e., process time allocation $t_c/t_h$, compared to steady-state engines. Substituting Eq.~\eqref{gamma} into Eq.~\eqref{PYvonmax}, the maximum power of the CA engine~\cite{curzonEfficiencyCarnotEngine1975} is exactly recovered
\begin{equation}
    P_{\mathrm{max}}^{\mathrm{(CA)}}=\frac{\gamma_h \gamma_c T_h}{\xi \left( \sqrt{\gamma_h}+\sqrt{\gamma_c} \right)^2}\left(1-\sqrt{\frac{T_c}{T_h}}\right)^2.
\end{equation}
On the other hand, by combining $\gamma_{h,c}$ and $t_{h,c}$ into $\Gamma_{h,c}$ through Eq.~\eqref{Gamma}, the power of the extended Yvon engine given in Eq.~\eqref{PYvon} can be derived from Eqs.~(\ref{eq-power-CA}-\ref{eq-entropy-variation-CA}) of the CA engine. In this sense, the extended Yvon engine and the CA engine are equivalent.

Herein, we would like to elucidate why the optimization of the extended Yvon engine exhibits greater succinctness than that of the original CA engine. In the optimization process from Eq.~\eqref{PYvon} to Eq.~\eqref{PYvonmax}, the ratio of $\theta_c/\theta_h$ emerges as a unified quantity, decoupling its role in maximizing $P$ from the heat transfer coefficients $\Gamma_{h,c}$. By contrast, in Ref.~\cite{curzonEfficiencyCarnotEngine1975}, the authors individually calculated the derivatives of $P$ (Eq.~(\ref{eq-power-CA})) with respect to $\theta_c$ and $\theta_h$, subsequently determining $\theta^*_c$ and $\theta^*_h$ via $\partial P/\partial\theta_c=\partial P/\partial\theta_h=0$. Eventually, Curzon and Ahlborn arrived at Eq.~\eqref{optimalT} and ascertained its independence from $\Gamma_{h,c}$ or $\gamma_{h,c}$. Essentially, our derivation is more intuitive and effectively highlights the significant universality that the EMP of the endoreversible Carnot engine is invariant to the heat transfer coefficients. Nevertheless, $\Gamma_{h,c}$ impact the specific values of $\theta^*_c$ and $\theta^*_h$ via Eqs.~\eqref{relation} and \eqref{optimalT}. Furthermore, when correlating heat transfer coefficients in the extended Yvon engine with that of the CA heat engine, Eq.~\eqref{gamma} indicates that the time ratio $t_c/t_h$ of different thermodynamic processes in the cyclic heat engine affects the output power as a unified quantity rather than depending on the specific durations of each individual process.

\section{\label{Sec:conclusion}Conclusion}
This work extends the original model of the Yvon engine with one-sided heat flux to the general endoreversible Carnot engine with two-sided heat fluxes. The extended Yvon engine and the CA engine are essentially two sides of the same coin, namely the steady-state heat engine form and the cyclic heat engine form of the endoreversible Carnot heat engine. Our study reveals why the EMP of this type of engine is independent of the heat transfer coefficients on both sides, emphasizing the importance of the temperature ratio over specific temperature values in the optimization of general endoreversible engines satisfying Newtonian heat transfer. For endoreversible heat engines operating under heat transfer laws other than Newton's law, the EMP generally depends significantly on the heat transfer coefficients~\cite{chenEffectHeatTransfer1989}. The optimization approach presented in the current work is intuitive and straightforward. It provides pedagogical value for constructing a clear physical picture and may also inspire studies on optimizing both cyclic and steady-state heat engines. 

Our work highlights that the temperature ratio $\theta_c/\theta_h$ (or efficiency $\eta=1-\theta_c/\theta_h$) is the sole degree of freedom for steady-state endoreversible engines, independent of the heat transfer coefficients [see Eq.~(\ref{PYvon}) or (\ref{Petarelation})], and decouples from time allocation in cyclic engines [see Eq.~(\ref{gamma})]. It is worth mentioning that i) Ref.~\cite{reyes-ramirezGlobalStabilityAnalysis2014} also recognizes the independent role of efficiency, but only for the special case with symmetric heat transfer coefficients; ii) Bejan~\cite{bejanModelsPowerPlants1996} noticed the fact that different irreversible heat engine models, namely Chambadal's~\cite{chambadalCentralesNucleaires1957}, Novikov's~\cite{novikovEfficiencyAtomicPower1957} and Curzon and Ahlborn's~\cite{curzonEfficiencyCarnotEngine1975} share the same EMP, and he explained this with the theory of entropy generation minimization~\cite{bejanEntropyGenerationMinimization1996a}. Nevertheless, the analyses presented in Ref.~\cite{bejanModelsPowerPlants1996} did not clarify the strict correspondence between the optimization of cyclic heat engines, which incorporated the process durations as per Curzon and Ahlborn's methodology, and that of steady-state heat engines characterized by energy fluxes~\cite{yvonSaclayReactorAcquired1955,chambadalCentralesNucleaires1957,novikovEfficiencyAtomicPower1957}. We provide such strict correspondence relations in Sec.~\ref{Sec:comparison} to fill this knowledge gap. 

As a final remark, perhaps due to historical, linguistic, or other factors~\cite{moreauCarnotPrincipleIts2015,ouerdaneContinuityBoundaryConditions2015}, as well as the particularity of the model, the Yvon engine~\cite{yvonSaclayReactorAcquired1955} and its contemporaneous works~\cite{chambadalCentralesNucleaires1957,novikovEfficiencyAtomicPower1957} regrettably did not garner the attention they deserved at that time. Twenty years later, the generality and simplicity of the CA engine~\cite{curzonEfficiencyCarnotEngine1975}, along with the systematic research on thermodynamics in finite time by the Chicago school~\cite{andresenThermodynamicsFiniteTime1977,berryFiniteTimeThermodynamics2022} during the same period, gave birth to the field of finite-time thermodynamics. We hope that the extended Yvon engine proposed in this paper will enable more people to appreciate Yvon's ingenious ideas regarding the practical thermodynamic cycle within a finite duration and help to disseminate the complete history of finite-time thermodynamics.

\begin{acknowledgments}
The authors thank Ruo-Xun Zhai for valuable comments on a preliminary version of this manuscript. This work is supported by the National Natural Science Foundation of China under grant No. 12305037 and the Fundamental Research Funds for the Central Universities under grant No. 2023NTST017.
\end{acknowledgments}

\bibliography{ref}
\end{document}